# Structural, optical and magnetic properties of nanostructured Cr-substituted Ni-Zn spinel ferrites synthesized by a microwave combustion method

Abdulaziz Abu El-Fadl[1,**], Azza M. Hassan[1], Mohamed A. Kassem[1,2,†]

1 *Department of Physics, Faculty of Science, Assiut University, Assiut 71516, Egypt,*

2 *Department of Materials Science and Engineering, Kyoto University, Kyoto 606-8501, Japan*

**Abstract**

Nanoparticles of $Cr^{3+}$-substituted Ni-Zn ferrites with a general formula $Ni_{0.4}Zn_{0.6-x}Cr_xFe_2O_4$ ($x$ = 0.0 - 0.6) have been synthesized via a facile microwave combustion route. The crystalline phase has been characterized by using X-ray diffraction (XRD), transmission electron microscopy (TEM), Fourier transform infrared spectroscopy (FT-IR) and X-ray photoelectron spectroscopy (XPS) revealing the spinel ferrite structure without extra phases. Crystallite sizes of 23 - 32 nm as estimated by XRD analyses, after corrections for crystal stains by Williamson–Hall method, are comparable to the average particle sizes observed by TEM which indicates successfully synthesized nanocrystals. Rietveld refinement analyses of the XRD patterns have inferred a monotonic decrease behavior of the lattice parameter with Cr doping in agreement with Vegard's law of solid solution series. Furthermore, cations distribution with an increased inversion factor indicate the B-site preference of $Cr^{3+}$ ions. The oxidation states and cations distribution indicated by XPS results imply the $Cr^{3+}$ doping on the account of $Zn^{2+}$ ions and a partial reduction of $Fe^{3+}$ to $Fe^{2+}$ to keep the charge balance in a composition series of $(Ni^{2+})_{0.4}(Zn^{2+}, Cr^{3+})_{0.6}(Fe^{2+}, Fe^{3+})_2(O^{2-})_4$. The optical properties were explored by optical UV-Vis spectroscopy indicating allowed direct transitions with band gap energy that decreases from 3.9 eV to 3.7 eV with Cr doping. Furthermore, the photocatalytic activity for the degradation of methyl orange (MO) dye was investigated showing largely enhanced photodecomposition up to 30 % of MO dye over $Ni_{0.4}Cr_{0.6}Fe_2O_4$ for 6 hours. A vibrating sample magnetometry (VSM) measurements at room temperature show further enhancement in the saturation magnetization of $Ni_{0.4}Zn_{0.6}Fe_2O_4$, the highest in Ni-Zn ferrites, from about 60 to 70 emu/g with the increase of Cr concentration up to $x$ = 0.1, while the coercivity shows a general increase in the whole range of Cr doping.

**Key words:** Microwave combustion method, spinel ferrites, XRD, FTIR, TEM, VSM

Corresponding authors:

† M. A. Kassem, email: makassem@aun.edu.eg

* A. A. El-Fadl , email: abuelfadl@aun.edu.eg, abulfadla@yahoo.com

# 1. Introduction

Magnetic spinels, $AB_2O_4$ with $A$ and $B$ are divalent and/or trivalent transition metals, have attracted much interest particularly in nanosized forms because of their fascinating magnetic, optical and electrical properties with theoretical and technological values [1–5]. Among spinel compounds, spinel ferrites ($AFe_2O_4$) have most remarkable feature of their physical and chemical properties that can also be tuned via substitutions by a divalent or a trivalent transition metal cations at the tetrahedral and octahedral sites [6–11]. A subsequent modified cation distribution after substitutions plays an important role in tailoring their magnetic and optical behavior [8]. Among various spinel ferrites, the antiferromagnetic (AF) zinc ferrite, $ZnFe_2O_4$, shows striking changes in its magnetic properties by reducing the grain size to the nanometer-sized range [12,13]. It is well known that bulk zinc ferrite is a normal spinel structure with Zn ions in the tetrahedral (A-sites) and Fe ions in the octahedral (B-sites) [14]. As $Fe^{3+}$ ions form a pyrochlore networks in the cubic spinel structure, bulk Zn-ferrite has become a model to study the fundamental magnetic frustration. However, Zn-ferrite nanoparticles show ferromagnetic and superparamagnetic order [15]. On another hand, Ni ferrite is soft ferrimagnetic material which has an inverse spinel structure in bulk scale [16]. Mixed Ni-Zn ferrites nanoparticles are the subject of intensive studies [17,18], due to their relation to numerous technological applications [19].

Many researchers have been employed the divalent and trivalent metal ions substitution to upgrade the electrical, optical and magnetic properties of Ni-Zn ferrites nanocrystals [20–27]. The effects of replacing $Fe^{3+}$ by $Cr^{3+}$ on the physiochemical properties of the Ni–Zn ferrite is one of the most common substitution have been reported [27–29]. It was revealed that chromium substitution modified the physical properties of spinel ferrites but does not alter the spinel structure. Furthermore, it is well known that selecting a synthesis route plays a vital role in the physical characteristics of spinel ferrites. Mixed Ni–Zn ferrites have been produced by numerous synthesis methods such as sol-gel [30], oxalate co-precipitation [31], hydrothermal technique [32]. Recently, Ni-Zn ferrites nanoparticles has been synthesized within short time by a simple and cost-effective microwave combustion method [33]. A. Abu El-Fadl et al. [10] have reported the structural and magnetic properties of Ni-Zn ferrites synthesized by microwave combustion and found that the composition $Ni_{0.4}Zn_{0.6}Fe_2O_4$ ferrite exhibits optimum magnetic properties. To our knowledge, no much work is done on replacing the divalent ions in spinel materials with trivalent cations, for instance, $Cr^{3+}$-substitution in $Ni_{1-y}Zn_{y-x}Cr_xFe_2O_4,$ such doping is absence from literature.

As a result, the present study focuses on the easy synthesis of $Ni_{0.4}Zn_{0.6}$ ferrite and study of their physical properties. The prepared ferrites were characterized using

X-ray diffraction (XRD), Fourier transform infrared spectroscopy (FT-IR) and transmission electron microscopy (TEM) measurements. Ultraviolet–visible (UV–Vis) spectroscopy measurements are utilized to study the optical properties. The dependences of the optical energy gap, photocatalytic activity and magnetic properties on the concentration of $Cr^{3+}$ ions are studied in detail.

## 2. Experimental techniques

*2.1. Materials and synthesis*

Spinel ferrite nanoparticles with the composition $Ni_{0.4}Zn_{0.6-x}Cr_xFe_2O_4$, $x$=0.0-0.6 were prepared in a step of 0.1 using a microwave combustion route. In a stoichiometric ratio, analytical grade metal nitrates: $Zn(NO_3)_2.6H_2O$, $Ni(NO_3)_2.6H_2O$, $Fe(NO_3)_3.9H_2O$ and $Cr(NO_3)_3.6H_2O$ were dissolved with glycine $((NH_2)_2COOH)$ as a fuel in small amount of distilled water by a magnetic stirrer. The produced solution was introduced for 20 min into a microwave oven (Olympic electric, KOR-6Q1B) operating at maximum power of 800 W. A brown-to-black voluminous and fluffy product was produced that were ground into fine powders.

*2.2. Characterization and measurements*

Phase identification and structural characterization of the prepared samples were carried out first by powder *X-ray diffraction* (XRD) using a diffractometer equipped in Bragg-Brentano geometry with an automatic divergent slit (Philips PW1710, Netherlands) uses CuKα-radiation of wavelength, $\lambda$= 0.15418 nm. The particle size and their morphologies were characterized using high resolution *transmission electron microscopy* (HR-TEM). Further characterization by Fourier transform infrared spectroscopy was carried out in the range 400-4000 cm$^{-1}$ by using *FT-IR spectrophotometer* (470 Shimadzu, 400-4000 cm$^{-1}$) by employing the KBr pellet method. *X-ray photoelectron spectroscopy* (XPS) was used to study the phase structure and chemical oxidation states of the synthesized samples. Powder samples have been compressed in pellets shape with highly flat surface and the collected XPS patterns have been corrected for C 1s signal at 285 eV. UV-Visible optical absorbance spectra have been collected from a suspension of the powder samples prepared by adding 10 mg of the sample to 10 mL of a DMSO solvent using a Thermo Evolution 300 *UV-Visible spectrophotometer* in a wavelength range of 200 -900 nm. *Photocatalytic activities* for methyl orange (MO) degradation have been studied by measuring the dye UV-Visible optical absorbance spectra after different times of photodegradation. 50 mg of nanoparticles were dispersed into 200 ml of freshly prepared solution of methylene orange (~1.65 ppm). This solution was illuminated by UV irradiation source. Magnetization curves were measured at room temperature by using a Lakeshore-7400 Series *vibrating sample magnetometer* (VSM).

## 3. Results and discussion
### 3.1. X-ray diffraction and structural properties

The X-ray diffraction patterns for the $Ni_{0.4}Zn_{0.6-x}Cr_xFe_2O_4$ series are shown in Fig. 1(a). The diffraction patterns exhibit a crystalline nature with all possible reflections belong to a spinel ferrite phase, indexed in the figure, without no impurity peaks were detected. The line broadening of diffraction peaks is attributed to the nanocrystalline nature and partial contribution of the internal strain. Further, the peaks show shift position to higher angles, as seen in Fig. 1(b), which indicates lattice shrinkage with increasing the Cr content.

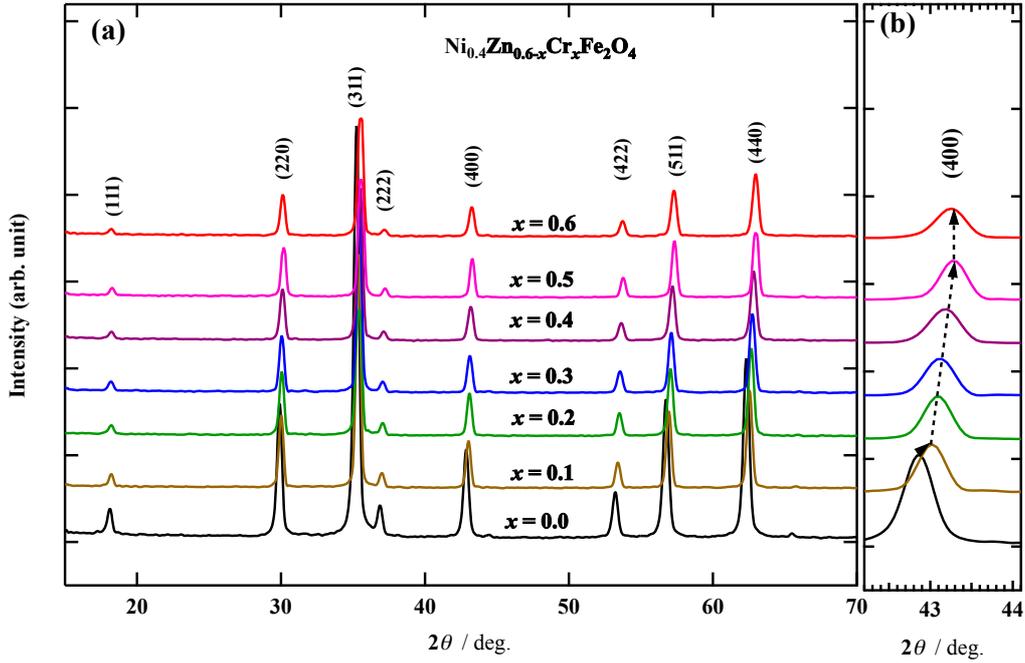

Figure 1. XRD patterns of $Ni_{0.4}Zn_{0.6-x}Cr_xFe_2O_4$ nanoparticles.

We have performed Rietveld refinement to accurately estimate the structural parameters as well as the cations distribution in $Ni_{0.4}Zn_{0.6-x}Cr_xFe_2O_4$ nanoparticles using the RIETAN-FP system for pattern-fitting structure refinement [34]. Figure 2 shows the XRD Rietveld refinement results for selected samples with $x$ = 0.0, 0.2, 0.4 and 0.6. The XRD patterns were calculated during refinements analysis by assuming the cubic $Fd\bar{3}m$ spinel structure with Wyckoff sites 8a and 16d for atoms in the tetrahedral A- and octahedral B-sites with fixed coordinates (1/8, 1/8, 1/8) and (1/2, 1/2, 1/2), respectively, while the site 32e with refined O coordinates ($x = y = z = u$). It is well known that both Ni and Cr ions prefer the octahedral B site with large stabilization energy in structures of the inverse spinel ferrite $NiFe_2O_4$ and all normal spinel chromites, respectively, while Zn can distribute between both A and B sites in nano spinel ferrites[10,35–37]. Our trials, indeed, have confirmed the best Rietveld refinement for the two end members, $Ni_{0.4}Zn_{0.6}Fe_2O_4$ and $Ni_{0.4}Zn_{0.6}Fe_2O_4$, when we consider these preferences. We have assumed the occupation of Ni and Cr ions in the octahedral B sites for the whole series XRD refinements. The lattice constant ($a$), oxygen position ($u$), occupancies of atoms ($g$) and peaks shape parameters such as FWHM, position, intensity, etc. and atomic displacements were refined during the Rietveld analysis. Structure parameters including lattice constant ($a$), oxygen position ($u$), X-ray density ($d_{XRD}$), crystallite size ($D_{XRD}$), lattice strain ($\varepsilon$) derived from the Rietveld analysis are summarized in Table 1.

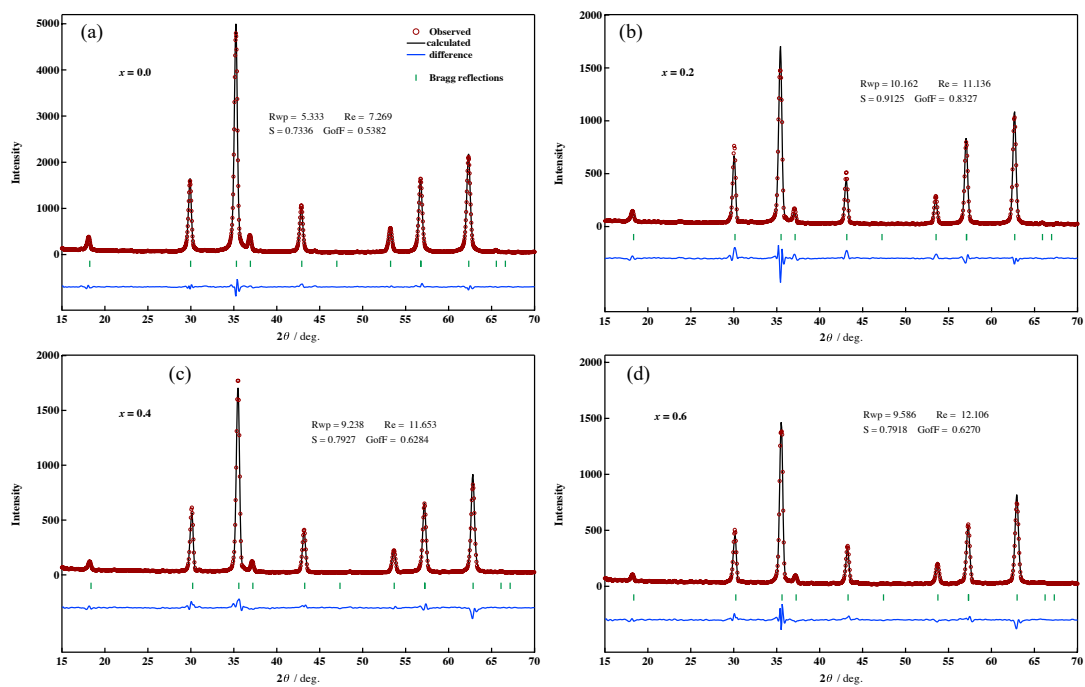

Figure 2. Rietveld analysis of XRD patterns of $Ni_{0.4}Zn_{0.6-x}Cr_xFe_2O_4$ nanoparticles

The Cr-content dependence of obtained values of $a$ is shown in Fig. 3(a). It is noticeable from the figure that Cr incorporation decreases the lattice constant first largely from 8.428 to 8.398 Å for $x$ = 0.0 – 0.1 then in a monotonic decrease behavior to a = 8.351 Å for $x$ = 0.5 and finally little decreased by Zn disappearance. The behavior of $a$ vs $x$ is consistent with the peaks shift in Fig. 1(b) and in a good agreement with Vegard's law [38], indicating successful substitution. The decrease of lattice constant with increasing $Cr^{3+}$ concentration is attributed to the smaller size of substituent $Cr^{3+}$ ions (0.615 Å in radius) relative to the replaced $Zn^{2+}$ ions (0.074 Å) in the B-site octahedral coordination [39]. The refinement also results in an oxygen position $u$ with a decrease from 0.381 to 0.378 with Cr doping in the whole series.

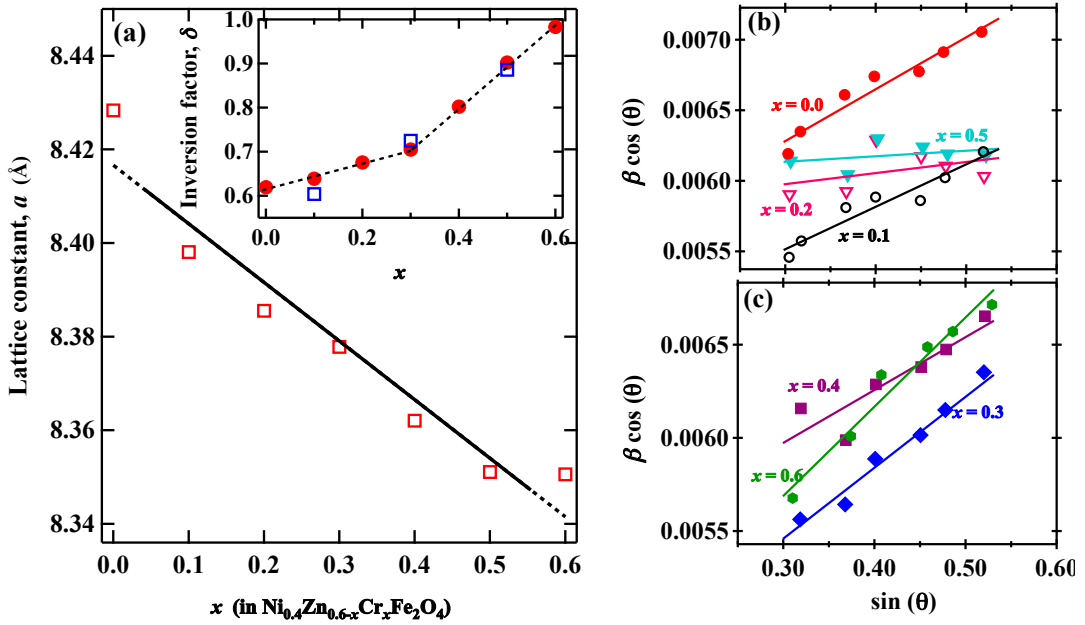

Figure 3. (a) Lattice constant and inversion factor (inset) as functions of Cr content, $x$, and (b) W-H analysis plots for $Ni_{0.4}Zn_{0.6-x}Cr_xFe_2O_4$ nanoparticles

The average crystallite sizes $D$ of the prepared spinel ferrites nanoparticles are estimated by using Williamson–Hall method [40]:

$$\beta\cos\theta = 4\varepsilon\sin\theta + \frac{k\lambda}{D}, \qquad (1)$$

Where $k$ is a constant which depends on the shape of the particle and almost equals 0.9, $\lambda$ is the X-ray wavelength, $\beta$ is the peak full width at half maximum (FWHM), $\theta$ is the diffraction angle and $\varepsilon$ is the internal strain of the lattice. equation (1) represents a linear plot of $\beta\cos\theta$ against $\sin\theta$ from which the internal strain ($\varepsilon$) and crystallite size ($D_{XRD}$) can be obtained from the slope and intercept with $y$-axis, respectively. Figures 3(b) and (c) show the $\beta\cos\theta$ against $\sin\theta$ plots of the synthesized compositions and the estimated $\varepsilon$ and are presented in table 1. The obtained crystallite sizes values of the prepared nanocrystals are in the range of 23–32 nm, and slight positive strain is observed which becomes almost negligible in the lattice of $x = 0.2$ and 0.5 samples.

The theoretical density can be calculated by using the relation [41]: $d_{XRD} = \frac{ZM}{N_A a^3}$, where $M$ is the molecular weight, $N$ is Avogadro's number, $Z$ is the number of atoms per unit cell and $a$ is the lattice constant. The calculated values of the X-ray density are given in table 1. The X–ray density increases with increasing concentration of Cr up to $x \simeq 0.1$ and then it starts to decrease with further Cr incorporation. This behavior can be explained by the competitive decrease in both *M* and *a* values with increasing Cr content. Since the lattice constant is drastically decreased with low Cr doping level and then slightly with further doping as seen in Fig. 3(a) while M decreases monotonically, thereby, the ratio ($M/a^3$) shows fluctuation behavior with $Cr^{3+}$ substitution.

**Table 1:** Structural parameters of $Ni_{0.4}Zn_{0.6-x}Cr_xFe_2O_4$ nanoparticles based on XRD analyses (Rietveld and Williamson–Hall) and TEM microscopy.

| *x* | *a* (Å) | *u* | $\varepsilon$ [± 0.0001] | $D_{XRD}$ [± 2] (nm) | $D_{TEM}$ [± 5] (nm) | $d_{XRD}$ (g/cm3) |
|---|---|---|---|---|---|---|
| 0 | 8.428 | 0.3805 | 0.00095 | 27.0 | | 5.2895 |
| 0.1 | 8.398 | 0.3795 | 0.00068 | 29.2 | | 5.3169 |
| 0.2 | 8.386 | 0.3799 | 0.00020 | 24.1 | 26 | 5.3106 |
| 0.3 | 8.378 | 0.3798 | 0.00095 | 32.1 | | 5.2950 |
| 0.4 | 8.362 | 0.3788 | 0.00061 | 26.3 | 23.20 | 5.2946 |
| 0.5 | 8.351 | 0.3779 | 0.00010 | 23.0 | | 5.2849 |
| 0.6 | 8.351 | 0.3776 | 0.00120 | 32.6 | 33.40 | 5.2553 |

The cations distribution estimated from XRD analysis is dependent on the Cr content indicated by the inversion factor, $\delta$, that increases with *x* as shown in the inset of Fig. 3(a). The observed non-monotonic behavior of $\delta$ showing a kink at *x* = 0.3 implies sites preferences. The first relatively slow increase in the inversion of $Ni_{0.4}Zn_{0.6}Fe_2O_4$ from 0.62 with Cr doping is explained by $Cr^{3+}$ substitution for $Zn^{2+}$ in the B site with movement of both $Zn^{2+}$ ions and $Fe^{3+}$ ions from the B site to the A site until all $Zn^{2+}$ ions become in the A site at around *x* = 0.3. For further substitution levels, *x* > 0.3, $Cr^{3+}$ still prefers the B site resulting in movement of $Fe^{3+}$ ions to the A site resulting in higher inversion of the spinel structure. The detailed results of cations distribution and atomic positions in the unit cell are presented in table 2.

3.2. **Transmission Electron Microscopy (TEM)**

TEM imaging has been reemployed to visualize the shape, size and morphology of the synthesized spinel ferrites. Figure 4 shows TEM micrographs of $Ni_{0.4}Zn_{0.6–x}Cr_xFe_2O_4$ nanoparticles with selected compositions of *x* = 0.4 and 0.6. The particles are mostly having octahedral as well as cubic shapes reflecting the high high-quality nonocrystals. The particles size varies in the range 5-50 nm and its distribution is shown in the insets of Figs. 3 with averages of 23 to 33 nm matching or slightly greater than the estimated values form XRD patterns. It is also observed that, the particles are agglomerated to some extent due to the fast reaction of the microwave combustion method and due the evolution of large amount of gases during reaction.

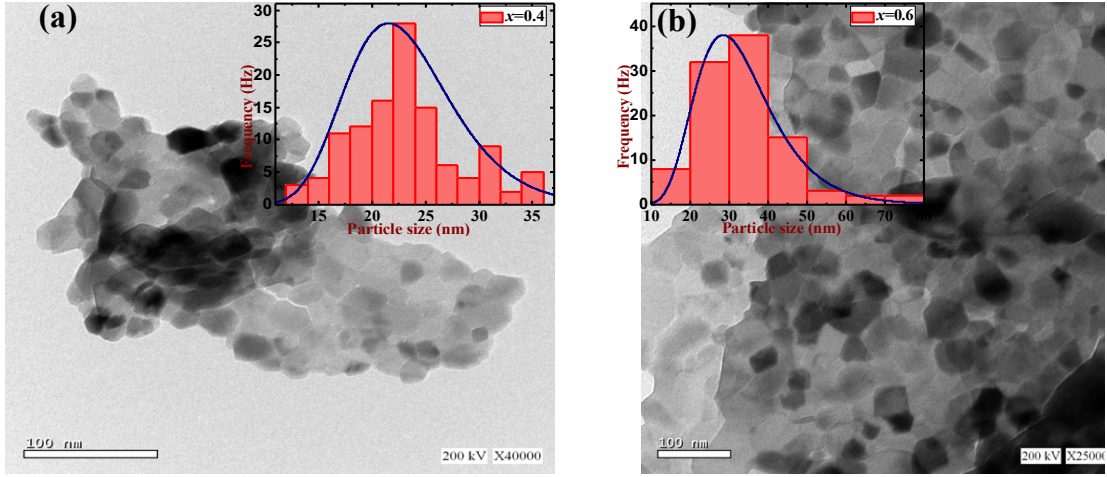

Figure 4. TEM images for Ni$_{0.4}$Zn$_{0.6-x}$Cr$_x$Fe$_2$O$_4$ nanoparticles with the Cr-contents of (a) $x = 0.4$ and (b) $x = 0.6$

Table 2: Cations distributions of Ni$_{0.4}$Zn$_{0.6-x}$Cr$_x$Fe$_2$O$_4$ nanoparticles by XRD Rietveld analysis.

| $x$ (Cr$^{3+}$) | Site (Wyckoff) | cations distribution | | Rietveld parameters | | |
|---|---|---|---|---|---|---|
| | | $x = y = z$ | Occupancy | $\delta$ | $S$ | $GofF$ |
| 0 | A$^{IV}$ (8a) | 0.125 | 0.381Zn + 0.619Fe | 0.619 | 0.734 | 0.538 |
| | B$^{VI}$ (16d) | 0.5 | 0.219Zn + 0.4Ni + 1.38Fe | | | |
| | O (32e) | 0.3805 | 4O | | | |
| 0.1 | A$^{IV}$ (8a) | 0.125 | 0.361Zn + 0.639Fe | 0.639 | 0.651 | 0.424 |
| | B$^{VI}$ (16d) | 0.5 | 0.139Zn + 0.4Ni + 0.1Cr + 1.361Fe | | | |
| | O (32e) | 0.3805 | 4O | | | |
| 0.2 | A$^{IV}$ (8a) | 0.125 | 0.324Zn + 0.676Fe | 0.676 | 0.913 | 0.833 |
| | B$^{VI}$ (16d) | 0.5 | 0.076Zn + 0.4Ni + 0.2Cr + 1.324Fe | | | |
| | O (32e) | 0.3805 | 4O | | | |
| 0.3 | A$^{IV}$ (8a) | 0.125 | 0.295Zn + 0.705Fe | 0.705 | 0.942 | 0.888 |
| | B$^{VI}$ (16d) | 0.5 | 0.005Zn + 0.4 Ni + 0.3Cr + 1.295Fe | | | |
| | O (32e) | 0.3805 | 4O | | | |
| 0.4 | A$^{IV}$ (8a) | 0.125 | 0.198Zn + 0.802Fe | 0.802 | 0.782 | 0.612 |
| | B$^{VI}$ (16d) | 0.5 | 0.002Zn + 0.4 Ni + 0.4Cr + 1.198Fe | | | |
| | O (32e) | 0.3805 | 4O | | | |
| 0.5 | A$^{IV}$ (8a) | 0.125 | 0.098Zn + 0.902Fe | 0.902 | 1.087 | 1.18 |
| | B$^{VI}$ (16d) | 0.5 | 0.002Zn + 0.4 Ni + 0.5Cr + 1.098Fe | | | |
| | O (32e) | 0.3805 | 4O | | | |
| 0.6 | A$^{IV}$ (8a) | 0.125 | 0.0169Ni + 0.983Fe | 0.983 | 0.7918 | 0.627 |
| | B$^{VI}$ (16d) | 0.5 | 0.383 Ni + 0.6Cr + 1.017Fe | | | |
| | O (32e) | 0.3805 | 4O | | | |

Moreover; this agglomeration behavior can be raised due to the permanent magnetic character of the present nanoferrite crystals. It is noticeable from the images that the degree of agglomeration between particles decreases with increasing Cr content, which can be attributed to the decrease in the net magnetic moment with further increase of Cr concentration.

### 3.1. X-ray photoelectron spectroscopy

As we have substituted with a trivalent ions $Cr^{3+}$ for a divalent $Zn^{2+}$, a partial reduction of some ionic species while the synthesis is expected to keep the charge neutrality. We have employed the XPS spectroscopy to investigate for the oxidation states and further for the cations distribution. Figure 5(a) shows a survey XPS scan for pellets samples with $x$ = 0.1, 0.3 and 0.5, from which the core photoionization signals of metals Zn 2p, Ni 2p, Fe 2p, Cr 2p, oxygen O 1s and hydrocarbon C 1s as well as and Auger signals of O KLL, Fe LMM, Ni LMM and Cr LMM are clearly displayed[42]. To investigate for the oxidation states, relative intensities and cations distributions, narrow scans have been collected over sufficiently long time for the Zn 2p, Ni 2p, Fe 2p, Cr 2p signals shown in Figs. 5(b-e). For $x$ = 0.1, the peaks binding-energy positions at 1021.8 and 1036.5 eV are corresponding to the Zn $2p_{3/2}$ and its shake-up satellite while the weaker peak at 1033.8 eV is for Zn $2p_{1/2}$ signal. The assignment of other metal core signals are as follows, Ni $2p_{3/2}$ (855.4 eV) and its satellite (861.8 eV), Ni $2p_{1/2}$ (873 eV) and its satellite (880.3 eV); Fe $2p_{3/2}$ (711.4 eV) and its satellite (719.3 eV), Fe $2p_{1/2}$ (725 eV) and its satellite (733.9 eV); Cr $2p_{3/2}$ (576.8 eV) and Cr $2p_{1/2}$ (586.8 eV) with hardly observed satellites. The most important confirmation result of XPS spectra is the gradual increase in the Cr content on the account of Zn rather than Fe nor Ni content clearly seen in figures change behavior of the signals intensity and shape. All peaks show slight position shift by Cr doping with almost unchanged peaks intensities for Fe 2p and Ni 2p peaks, Figs. 5(b) and (e), while gradual intensity decrease of Zn 2P and increase of Cr 2p peaks are observed, Figs. 5(c) and (d), with Fe to (Ni+Zn+Cr) relative atomic ratio of about 2. The peaks assignment indicated the most stable oxidation state of $Zn^{2+}$, $Ni^{2+}$ and $Cr^{3+}$ and dominant $Fe^{3+}$. To elucidate the cations distribution and Fe valance states, the deconvolution of the Fe 2p peaks, including only Fe $2p_{3/2}$ satellite for simple Shirley background shape, confirmed the presence of Fe ions in the two lattice sites, tetrahedral (A) and octahedral (B) sites, shown in Fig. 5(f) for $x$ = 0.3. The low binding-energy Fe $2p_{3/2}$($2p_{1/2}$) sub-peaks centered around 710.25 (723.64) eV are assigned to the $Fe^{2+}$ at B site while the sub-peaks centered around 711.3 (724.8) eV and 712.96 (727.12) eV are assigned to $Fe^{3+}$ ions at B and A sites, respectively[43]. The partial reduction of $Fe^{3+}$ to $Fe^{2+}$ is expected to occur while this microwave synthesis during quick reaction to compensate for the charge unbalance caused by $Cr^{3+}$ ions as previously observed in hexaferrite synthesis at high temperatures. The inversion factor, $\delta$, values estimated as the ratio of $Fe^{3+}$(A-site) relative to the total Fe content match values obtained by Rietveld analysis of the XRD data and are aslo shown in the inset of Fig. 3(a).

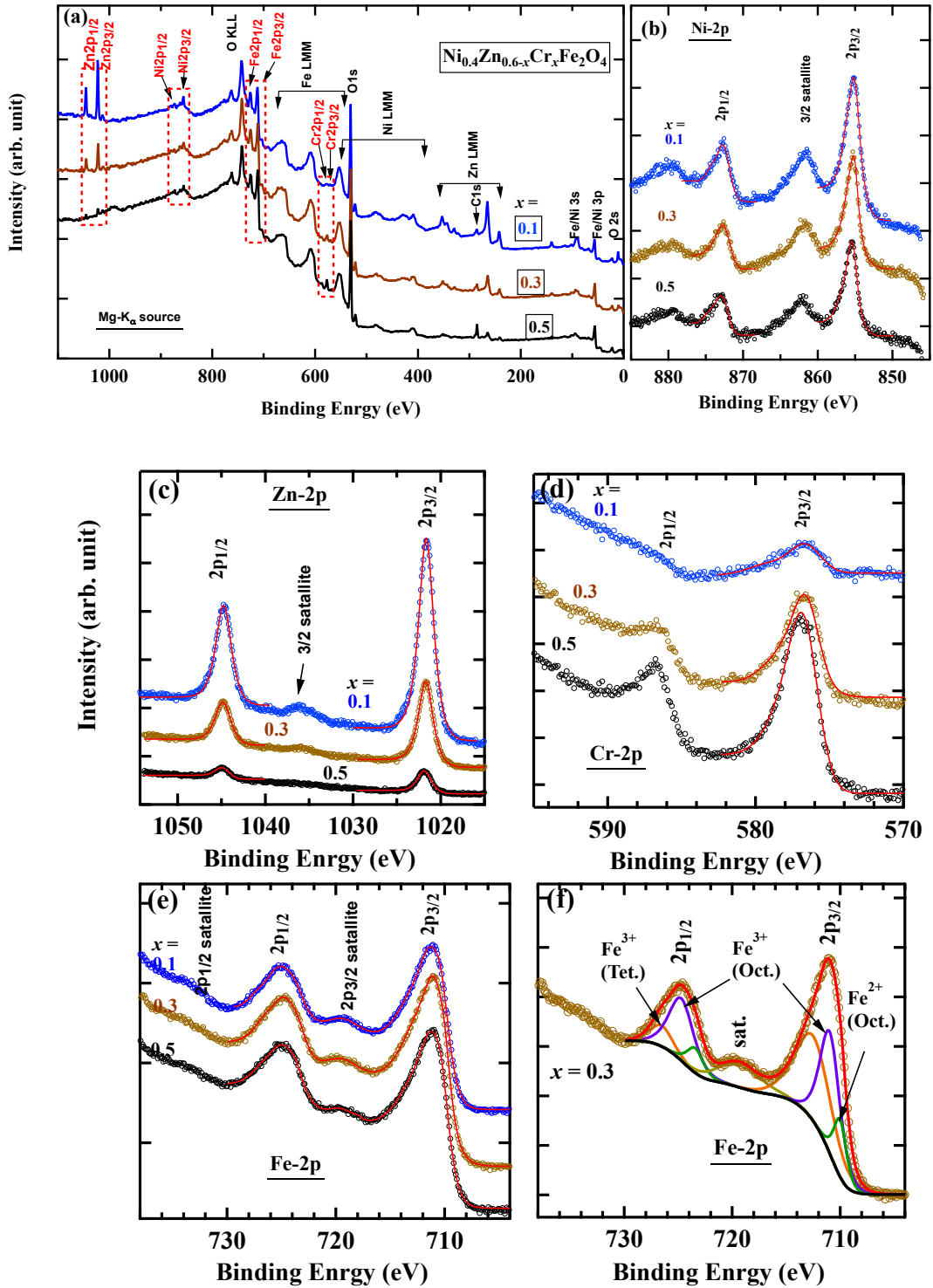

**Figure 5.** XPS spectra of $Ni_{0.4}Zn_{0.6-x}Cr_xFe_2O_4$ nanoparticles: (a) survey scans for $x$ = 0.1, 0.3 and 0.5, (b) – (e) long-time measured patterns of transition metal 2p bands and (f) an example of the Fe-2p band analysis.

### 3.2. Fourier Transform Infrared Spectroscopy (FTIR) Analysis

The FTIR spectra of $Ni_{0.4}Zn_{0.6-x}Cr_xFe_2O_4$ nanoparticles were measured in the frequency range 400 – 4000 cm$^{-1}$ and are shown in Fig. 6(a) below about 1300 cm$^{-1}$ for clarity. The spectra indicate the two characteristic absorption bands of spinel ferrites as reported by Waldron [44]. One band is completely observed at 565-583 cm$^{-1}$ while the onset of the other one which occure just below 400 cm$^{-1}$, lower limit of our measurements[9]. The high frequency band occurs at $\nu_1$ in the range 565 - 583 cm$^{-1}$ is related to the anti-symmetric stretching vibrations of the oxygen bond to metal ions in the A site, (Me-O)$_{Teth}$, and the lower frequency band, $\nu_2$, is about 360 cm$^{-1}$ is assigned to anti-symmetric stretching vibration of (Me-O)$_{Octh}$ bonds in B site[9]. This is because the octahedral site bond length is greater than that in the tetrahedral site and it is well known that the frequency band is inversely proportional to the bond length [45]. The $\nu_1$ band frequency values are presented in table 3. The value of $\nu_1$ shows an augment trend with increasing Cr content, which could be related to the substitution of relatively lighter Cr ions for Zn ions. It is well known that Cr ions have strong preference to occupy octahedral sites which also suggest the migration of some Fe ions to the tetrahedral sites [46,47].

### 3.3. Optical properties

The optical properties of $Ni_{0.4}Zn_{0.6-x}Cr_xFe_2O_4$ nanocrystals were investigated by UV-Visible spectrophotometry and the UV–Vis absorbance spectra are shown in Fig.6(b). The spectra show that the absorption edge is slightly shifted towards higher wavelength with increasing Cr concentration. The optical band gap $E_g$ can be estimated from the optical absorption spectra by using the well-known Tauc's relation [48]:

$$\alpha h\nu = \beta(h\nu - E_g)^n, \quad (4)$$

Where $\alpha$ is the absorption coefficient, $\nu$ is the frequency of the incident light, $h$ is Plank's constant and $n$ is a constant whose value determines the type of the electron transition. It can take the values 1/2, 2, 3, 3/2 for transition desired direct and indirect allowed transition, indirect forbidden and direct forbidden, respectively. Figs. 6(c) and (d) depicts the Tauc's plots of $Ni_{0.4}Zn_{0.6-x}Cr_xFe_2O_4$ nanocrystals at various substitution levels by Cr ions for direct allowed transitions. The intercept of the straight line on the *(hν)* axis corresponds to the direct allowed optical band gaps and estimated values are given in table 3. It is clearly shown from the table that the optical band gap of the synthesized nanoparticles are monotonically decreased from 3.9 eV to 3.78 eV with incorporating Cr ions by *x* from 0.1 to 0.6. The decrease in $E_g$ value with Cr ions doping is possibly a result of sub-band-gap energy levels formation [49]. Other factors such as crystallite size, lattice constant, agglomeration and lattice defects may have some effects on the $E_g$ [50,51]. However, the fine size effects are excluded due to the absence of change trends in the crystallite size, see table 1.

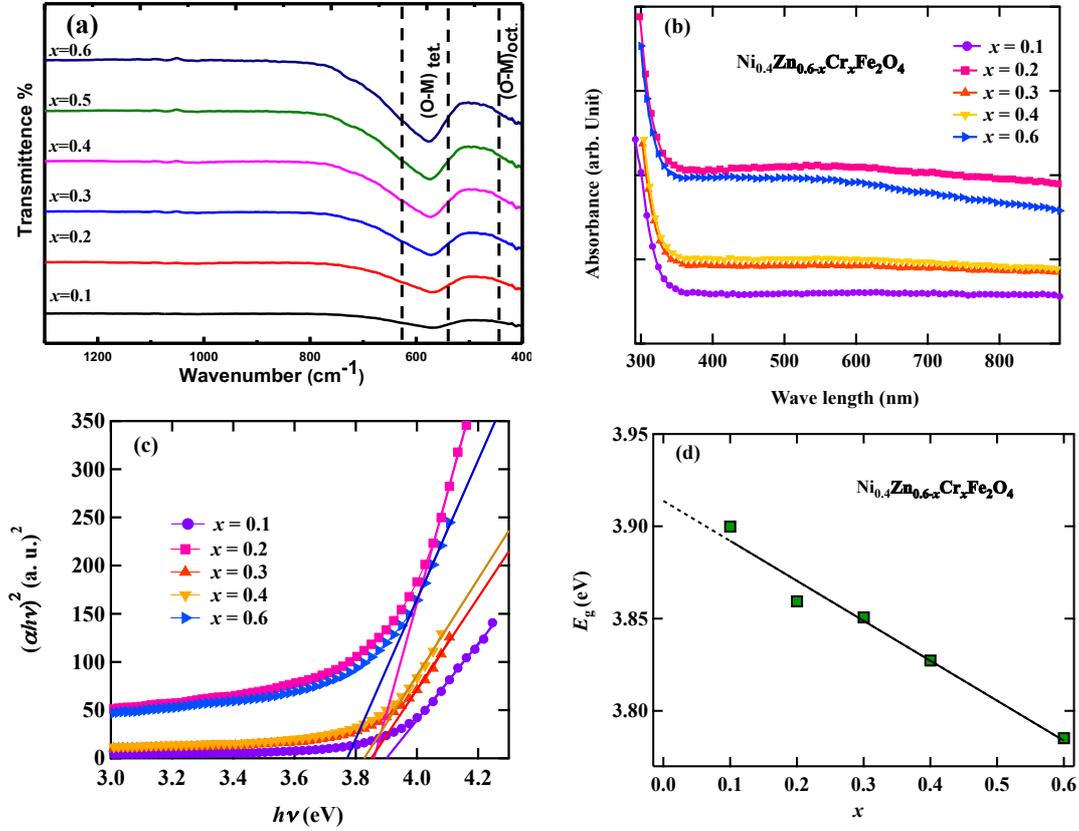

Figure 6: (a) FTIR- and (b) UV-visible absorbance spectra of $Ni_{0.4}Zn_{0.6-x}Cr_xFe_2O_4$ nanoparticles. (c) Tauc's plots of direct allowed transition and (d) the Cr-content dependence of the estimated direct energy gap, $E_g$.

**Table 3.** The observed optical energy gap values and FT-IR band frequencies of $Ni_{0.4}Zn_{0.6-x}Cr_xFe_2O_4$ nanocrystals.

| Cr-content (x) | $E_g$ (eV) | $\nu_1$ (cm$^{-1}$) |
|---|---|---|
| **0.1** | 3.90 | 565 |
| **0.2** | 3.86 | 570 |
| **0.3** | 3.85 | 572 |
| **0.4** | 3.83 | 578 |
| **0.5** | - | 579 |
| **0.6** | 3.78 | 580 |

### 3.4. Photocatalytic dye degradation studies

The photocatalytic activity of the synthesized $Ni_{0.4}Zn_{0.6-x}Cr_xFe_2O_4$ nanoparticles is evaluated for some selected compositions ($x$=0.0, 0.2, 0.4 and 0.6) for the degradation of methyl orange (MO) dye. Figure 7 shows the UV-visible absorbance spectra of MO dye degradation within the time interval of 0–360 min. It noticeable from the spectra figures that the maximum absorbance intensity, which is centered at 463 nm, decreases with increasing irradiation time in the presence of $Ni_{0.4}Zn_{0.6-x}Cr_xFe_2O_4$ photo-catalyst with irradiation by UV-visible light. Fig. 8(a) shows the photo degradation ratio of MO by $Ni_{0.4}Zn_{0.6-x}Cr_xFe_2O_4$ photocatalyst as function of the irradiation time. It is observed that the photocatalytic degradation efficiency was enhanced with increasing the concentration of Cr ions. This can be attributed to the observed decrease of the band gap energy from 3.89 eV to 3.78 eV. The percentage of dye removal shown in Fig. 8(b) was calculated using the following equation: $E = \frac{C_o - C}{C_o}$ %, where $C_0$ and $C$ represent the initial and final concentrations of the dye solution, respectively. From Fig. 8, it is observed that the highest value of photocatalytic activity, about 32 % after irradiation time of 6 hours, is achieved by the sample with $Ni_{0.4}Cr_{0.6}Fe_2O_4$ composition. The low activity in MO-dye degradation over the present nanoparticles in relatively to the reported data for $ZnFe_{2-x}Cr_xO_4$ nanoparticles can be explained by the main dependence of the photocatalytic effect on the band gap, as lower band gap in which the interaction of light with the material generates electron–hole pairs increases the photocatalytic activity. Other factors such as the particle size and porosity and hence surface area have high impact on the nanoparticles photocatalytic effect [52].

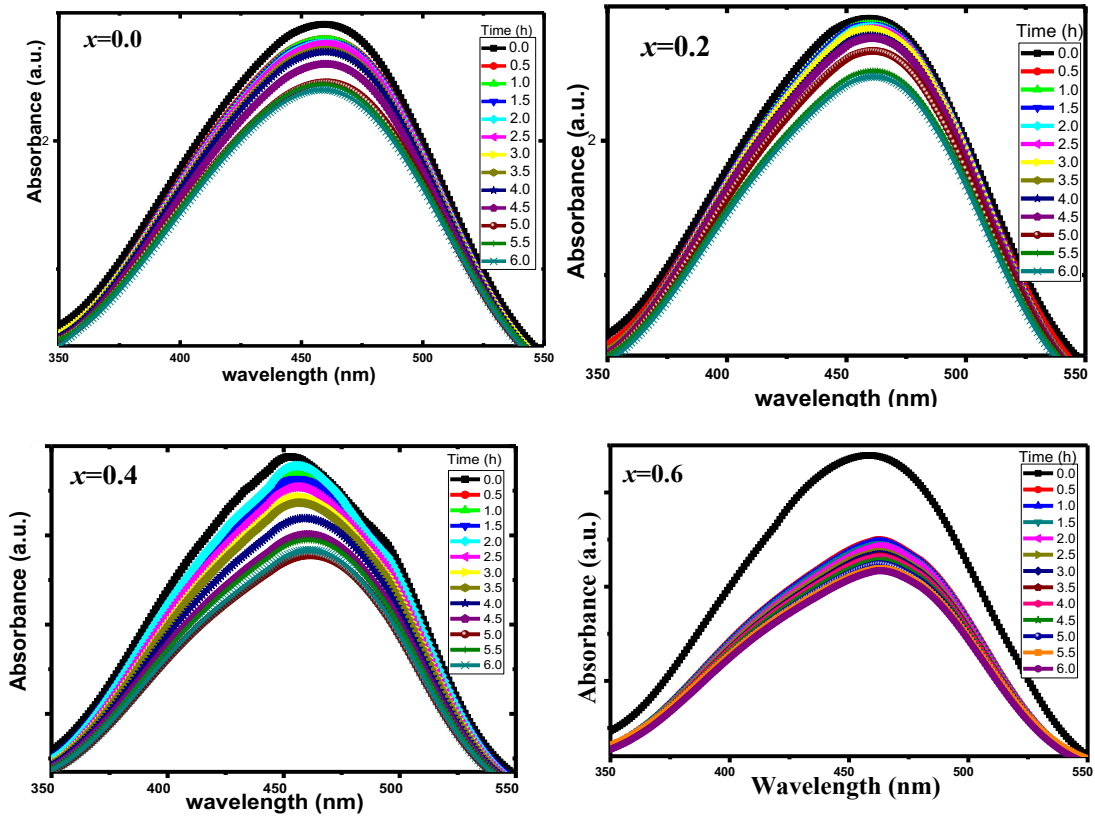

Figure 7. Variation of absorption spectra of orange I in presence of $Ni_{0.4}Zn_{0.6-x}Cr_xFe_2O_4$ nanoparticles under UV-visible light source.

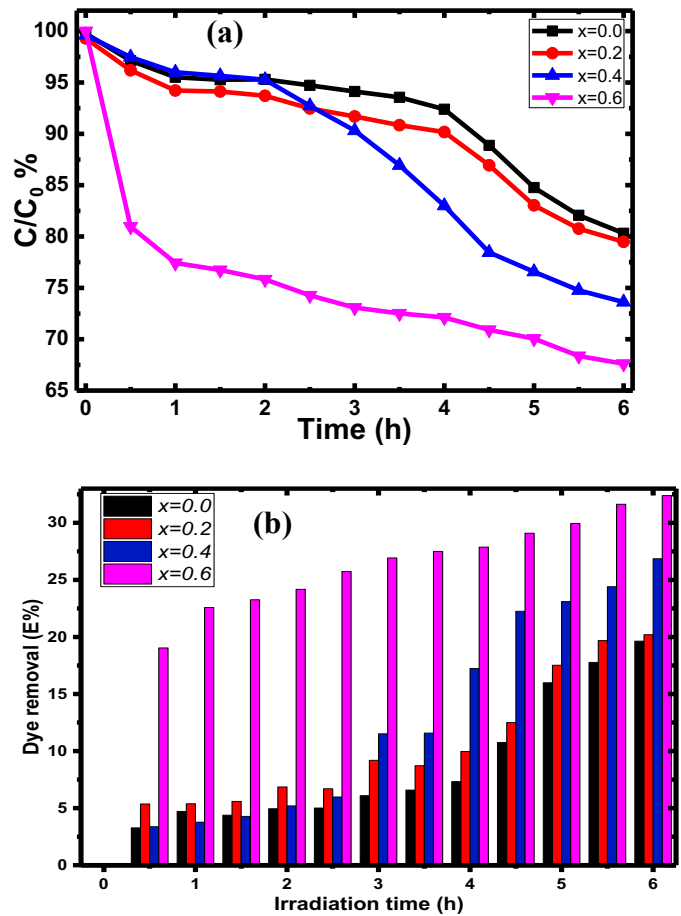

Figure 8. (a) relative concentrations (b) removal ration of MO dye over $Ni_{0.4}Zn_{0.6-x}Cr_xFe_2O_4$ nanoparticles

### 3.5. Magnetic studies

Figure 9(a) shows the magnetic hysteresis loops (*M-H* curves) for $Ni_{0.4}Zn_{0.6-x}Cr_xFe_2O_4$ nanocrystals measured at room temperature. It is clear from the figure that all samples exhibit a clear behavior of soft ferrimagnetic materials. The magnetic parameters such as saturation magnetization ($M_s$), remnant magnetization ($M_r$), coercivity ($H_c$) and anisotropy constant ($k = \frac{H_c M_s}{0.96}$) values are listed in table 4. Figure 9(b) shows the variation of $M_s$ and $H_c$ with the Cr-content. It is noticeable from the figure that the saturation magnetization $M_s$ increases with adding Cr ions up to $x =$ 0.1- 0.2 and then it decreases with further incorporation of Cr ions with ($x > 0.2$). The initial observed increase of the magnetization from 59.92 emu/g to 67.21 emu/g with addition of small amount of $Cr^{3+}$ can be discussed in terms of the cations redistribution between both A and B sites after Cr incorporation as well as the magnetic character of the constituent cations on both sites. It is well known that Cr and Ni ions have strong preference to occupy the octahedral site [53], thereby, the magnetic $Cr^{3+}$ ions (3 $\mu_B$) first replaces the nonmagnetic $Zn^{2+}$ ions (0 $\mu_B$) at B-site. Hence, the net magnetic moment is increased in according to the Néel's ferrimagnetic theory; the net magnetic moment can be represented as $\mu_s = \mu(B) - \mu(A)$, where $\mu(B)$ and $\mu(A)$ are the magnetization of B and A sites, respectively[54,55].

The value of $\mu(B)$ and hence $\mu_s$ become maximum when all Zn ions are moved to the A site at $x = 0.2$. At higher concentrations of Cr ions as a substituent of Zn ions the $M_s$ and $M_r$ values are decreased due to the decreased $\mu(B)$ with moving more Fe3+ ions (5 $\mu_B$) to the A site and replaced by $Cr^{3+}$ (3 $\mu_B$) and the emergent $Fe^{2+}$ ions (4 $\mu_B$). the magnetization results are in agreement with the cations distribution estimated based on XRD and XPS analyses. Similar behavior of $M_s$ was observed by Rostami and Majles et al. [56] As $Mg^{+2}$ ions are substituted by $Cu^{+2}$ ions in $Mg_{0.6}Ni_{0.4}Fe_2O_4$. Also Li et al. [57] have found that saturation magnetization of $Ni_{0.5-x}Zn_{0.5}Cr_xCo_{0.1}Fe_{1.9}O_4$ increased with the increase of Cr substitution when $x < 0.05$, and then decreased when $x > 0.05$. The coercivity exhibit a general modified trend with increasing Cr content which is mainly could be related to the higher magnetocrystalline anisotropy of Cr in compared to Zn and Fe ions and the variation of crystallite size of the synthesized nanocrystals [58].

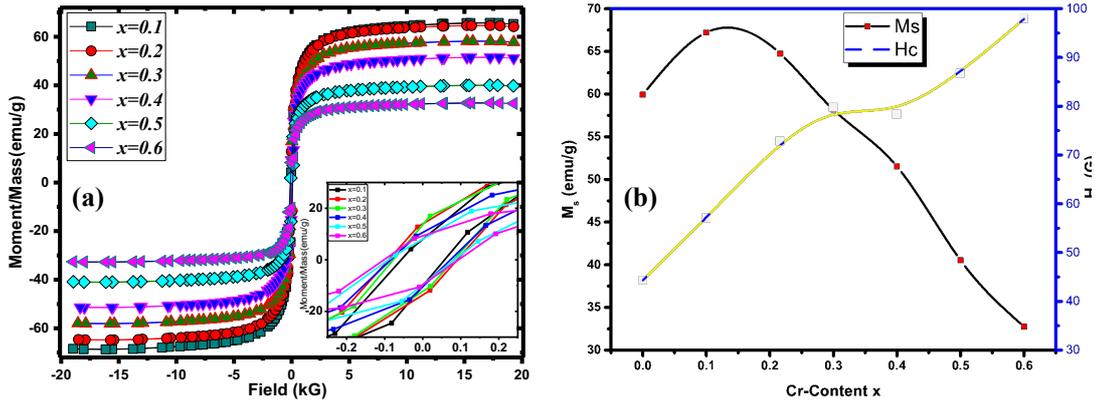

Figure 9: (a) Room temperature hysteresis loops and (b) variation of magnetic parameters $M_s$ and $H_c$ with the Cr content in $Ni_{0.4}Zn_{0.6-x}Cr_xFe_2O_4$ nanoparticles.

Table 4: Magnetic parameters of $Ni_{0.4}Zn_{0.6-x}Cr_xFe_2O_4$ nanoparticles

| $x$ [$Cr^{3+}$] | $M_s$ (emu/g) | $H_c$ (Oe) | $M_r$ (emu/g) | $K$ (ergs/cm$^2$) | $H_k$ (Oe) | $M$ ($\mu_B$) | R |
|---|---|---|---|---|---|---|---|
| **0.0** | 59.92 | 44.32 | 6.43 | 2709 | 90 | 2.55 | 107 |
| **0.1** | 67.21 | 57.08 | 9.05 | 3915 | 116 | 2.85 | 134 |
| **0.2** | 64.75 | 90.22 | 13.91 | 5960 | 184 | 2.73 | 214 |
| **0.3** | 58.16 | 79.69 | 12.92 | 4729 | 162 | 2.44 | 222 |
| **0.4** | 51.50 | 78.32 | 10.53 | 4116 | 159 | 2.14 | 204 |
| **0.5** | 40.53 | 86.77 | 8.84 | 3589 | 177 | 1.68 | 218 |
| **0.6** | 32.75 | 97.89 | 9.47 | 3272 | 199 | 1.35 | 289 |

## 4. Conclusion

Microwave combustion process was successfully used to fabricate $Ni_{0.4}Zn_{0.6-x}Cr_xFe_2O_4$ nanoparticles. XRD and FT-IR clearly exhibited the formation of single-phase spinel ferrite. The lattice parameters decreased with increasing Cr ion content owing to its smaller ionic radius in compared to Zn ion the octahedral coordination. The crystallite size varied from 20 nm to 30 nm. XPS analysis implies that the $Cr^{3+}$ doping is on the account of $Zn^{2+}$ ions and a partial reduction of $Fe^{3+}$ to $Fe^{2+}$ occurs which explains the charge balance after $Cr^{3+}$ doping. Cations distribution that explains the observed magnetic properties have been revealed by XRD-Rietveld and XPS analyses. UV–visible studies show that Cr ion doping caused significant decrease in the optical band gaps $E_g$, which is attributed to the formation of sub-levels among the energy band gaps. As a result, it was observed that photo catalytic activity of $Ni_{0.4}Zn_{0.6-x}Cr_xFe_2O_4$ nanoparticles or the degradation of MO is enhanced with increasing Cr ions substitutions. Magnetization measurement indicated that the saturation magnetization increased with the increase of Cr content up to about $x = 0.2$ and then it decreased with further Cr-doping, while the coercivity increases monotonically with Cr-doping.

**Acknowledgement**
Authors would like to acknowledge the help of Mr. Y Sonobayashi, from the Department of Materials Science and Engineering, Kyoto University, with measurements and great discussion of the XPS data.